\documentclass[twocolumn,superscriptaddress,floatfix,showpacs,aps]{revtex4}
\usepackage{amsmath,amssymb,epsfig,color}

\usepackage{graphicx}
\usepackage{dcolumn}
\usepackage{bm}
\usepackage{color}

\begin{document}

\title{Raman spectroscopy of the low dimensional antiferromagnet with large N\'eel temperature SrRu$_2$O$_6$.}
\pacs{}
\author{Yu.S. Ponosov}
\affiliation{M.N. Mikheev Institute of Metal Physics UB RAS, 620137, S. Kovalevskaya str. 18, Ekaterinburg, Russia}
\affiliation{Ural State Technical University, Mira St. 19, 620002 Ekaterinburg, Russia}

\author{E.V. Komleva}
\affiliation{M.N. Mikheev Institute of Metal Physics UB RAS, 620137, S. Kovalevskaya str. 18, Ekaterinburg, Russia}

\author{D.A. Zamyatin}
\affiliation{Zavaritsky Institute of Geology and Geochemistry UB RAS, 620016, Akademika Vonsovskogo str. 15, Ekaterinburg, Russia}

\author{R. I. Walton}
\affiliation {Department of Chemistry, University of Warwick, Gibbet Hill Road, Coventry, CV4 7AL, United Kingdom}

\author{S.V. Streltsov}
\affiliation{M.N. Mikheev Institute of Metal Physics UB RAS, 620137, S. Kovalevskaya str. 18, Ekaterinburg, Russia}
\affiliation{Ural State Technical University, Mira St. 19, 620002 Ekaterinburg, Russia}

\begin{abstract}
We report results of the Raman measurements for SrRu$_2$O$_6$ having extraordinary high N$\acute{\textrm{e}}$el temperature for a layered material. No additional phonon modes were detected at the temperature of magnetic transition thus excluding lowering of the symmetry in the magnetically ordered phase. An unusual increase in softening and damping of some phonons as the temperature approaches $T_N$ indicate the appearance of a continuum of interacting electronic excitations at $T\geq300K$. We also observe an intensive Raman response at 2050 cm$^{-1}$. Analysis of the polarization dependence and comparison with available theoretical data shows that this peak likely originates from the transitions between molecular orbitals previously proposed to explain the magnetic properties of SrRu$_2$O$_6$.
\end{abstract}    

\date{\today}

\maketitle

\section{Introduction}
Localization of the electrons not on the (site-centred) atomic, but on the (bond-centred) molecular orbitals often leads to rather unusual physical phenomena in transition metal oxides: various structural transitions\cite{Bulaevskii1975,Stucky1977}, different magnetic anomalies (including formation of the spin gap), orbital selective behaviour\cite{Streltsov-2016} and metal-insulator transitions\cite{Khomskii2014,Streltsov-2017}. Typically molecular orbitals can be found in materials, where isolated dimers, trimers or even more complex clusters of transition metals exist in a structure or where they are spontaneously formed by a structural phase transition. The situation, when there is no real structural clusters of transition metal ions, but a material under consideration still demonstrates some properties related to the formation of the molecular orbitals, is much more unique and much more interesting. It has been recently argued that this is exactly the case in SrRu$_2$O$_6$~\cite{st}.

The first report of the synthesis of SrRu$_2$O$_6$ was published in Ref.~\cite{Hiley2014}. The crystal structure  is layered and Ru ions form an ideal honeycomb lattice. SrRu$_2$O$_6$ is an antiferromagnetic (AFM) insulator (all nearest neighbors AFM ordered) with unexpectedly high N$\acute{\textrm{e}}$el temperature, $T_N \sim 565$~K\cite{Hiley2014,Hiley2015,Tian2015}. The magnetic moments are $\sim$1.3-1.4$\mu_B$,\cite{Hiley2015,Tian2015}, which is surprisingly small for the Ru$^{5+}$ ions with $4d^3$ electronic configuration having $S=3/2$. Magnetic susceptibility does not follow simple Curie-Weiss law in the paramagnetic region, increasing with elevated temperatures. This temperature behaviour is reminiscent of an itinerant magnet with a sharp feature in the electronic density of states, see e.g.~\cite{Shimizu1981,Moriya1984a,Streltsov2017b}.

Very different theoretical pictures have been proposed to explain the observed experimental features of SrRu$_2$O$_6$. The first one is based on the localized nature of $4d$ electrons and describes them as strongly correlated\cite{Tian2015}, stressing the orbital-selective behaviour of these electrons due to trigonal distortions of RuO$_6$ octahedra. In the dynamical meant field theory  calculations based on the density function theory band structure (DFT+DMFT) the $a_{1g}$ electrons appear to be localized, while $e_g^{\pi}$ are itinerant~\cite{Okamoto2017a}. This is similar to the double exchange picture typical for ferromagnetic interaction, but in a model presented in Ref.~\cite{Okamoto2017a} results in the N$\acute{\textrm{e}}$el AFM. Further DFT+DMFT calculations did not find the orbital-selective behaviour, but pointed out that the result strongly depends on the value of Hund's exchange $J_H$ and concluded that SrRu$_2$O$_6$ should be considered not as Mott-Hubbard, but rather as a covalent insulator\cite{Hariki2017}.

The fact that SrRu$_2$O$_6$ is the covalent insulator is consistent with an alternative (to purely Mott-Hubbard picture) model, which finds that the bare (uncorrelated) electronic spectrum can be described by the conception of molecular orbitals centred on Ru hexagons\cite{st}. The situation reminds benzene molecule. These molecular orbitals appear without any dimerization or clusterization, but are formed due to a specific symmetry of the $d-p-d$ electronic hopping on the honeycomb lattice. The electrons in this case turns out to be delocalized within the Ru hexagons, but the hoppings between the hexagons are suppressed. This conception explains why SrRu$_2$O$_6$ turns out to be a strongly non-Heisenberg magnet in the DFT calculations\cite{Singh2015,st} and predicts that the molecular orbitals may manifest themselves in experimental X-ray or optical spectra due to specific selection rules\cite{Pchelkina2016}. There are however factors such as, e.g., direct $d-d$ hopping, strong on-cite correlation effects, which tend to spoil molecular orbital picture.

While both Mott-Hubbard and molecular-orbitals pictures explain high N$\acute{\textrm{e}}$el temperature in a similar manner (due to interplanar exchange and single ion anisotropy) the electronic spectra in these models are different and further spectroscopical experimental studies are needed to distinguish between these two scenarios.

In this paper we present results of the Raman scattering measurements, which detect an additional (to phonon lines) peak at 2050 cm$^{-1}$. These excitations can be either due to two-magnon scattering or because of inter molecular orbital excitations. We find that frequency of this peak is inconsistent with available DFT estimations of exchange constants and thus conclude that its origin is rather connected with interband electronic transitions, but with the gap renormalized by the correlation effects. The symmetry of excitations is consistent with predictions based on the molecular orbitals picture.

\section{Experimental and calculation details}
Raman experiments were carried out on freshly cleaved surfaces of compacted disks of SrRu$_2$O$_6$ polycrystalline powder prepared hydrothermally at 473 K as described elsewhere ~\cite{Hiley2014}. The samples were characterized by variable temperature powder neutron diffraction  and magnetic susceptibility measurements performed on the Quantum Design MPMS-XL squid magnetometer with a furnace insert. 

Polarized Raman measurements  in the temperature range of 300 to 670 K were performed in backscattering geometry using RM1000 Renishaw microspectrometer equipped with 532 nm solid-state laser and 633 helium-neon laser. Low temperature measurements  in the temperature range of 80 to 360 K were performed using Ar-ion laser with wavelength 514 nm using Raman spectrometer Labram HR-800 coupled to a liquid-nitrogen–cooled CCD detector. Respective Linkam stages were used for temperature variation. The laser beam was focused (~5 $\mu$m in diameter) on microcrystals of hexagonal shape up to 20  $\mu$m in size (XY plane) or on thin rectangular crystals (XZ or YZ plane). Very low power (up to 0.1 mW) was used to avoid local heating of the crystals.
\begin{figure}[b]
\includegraphics[width=0.45\textwidth]{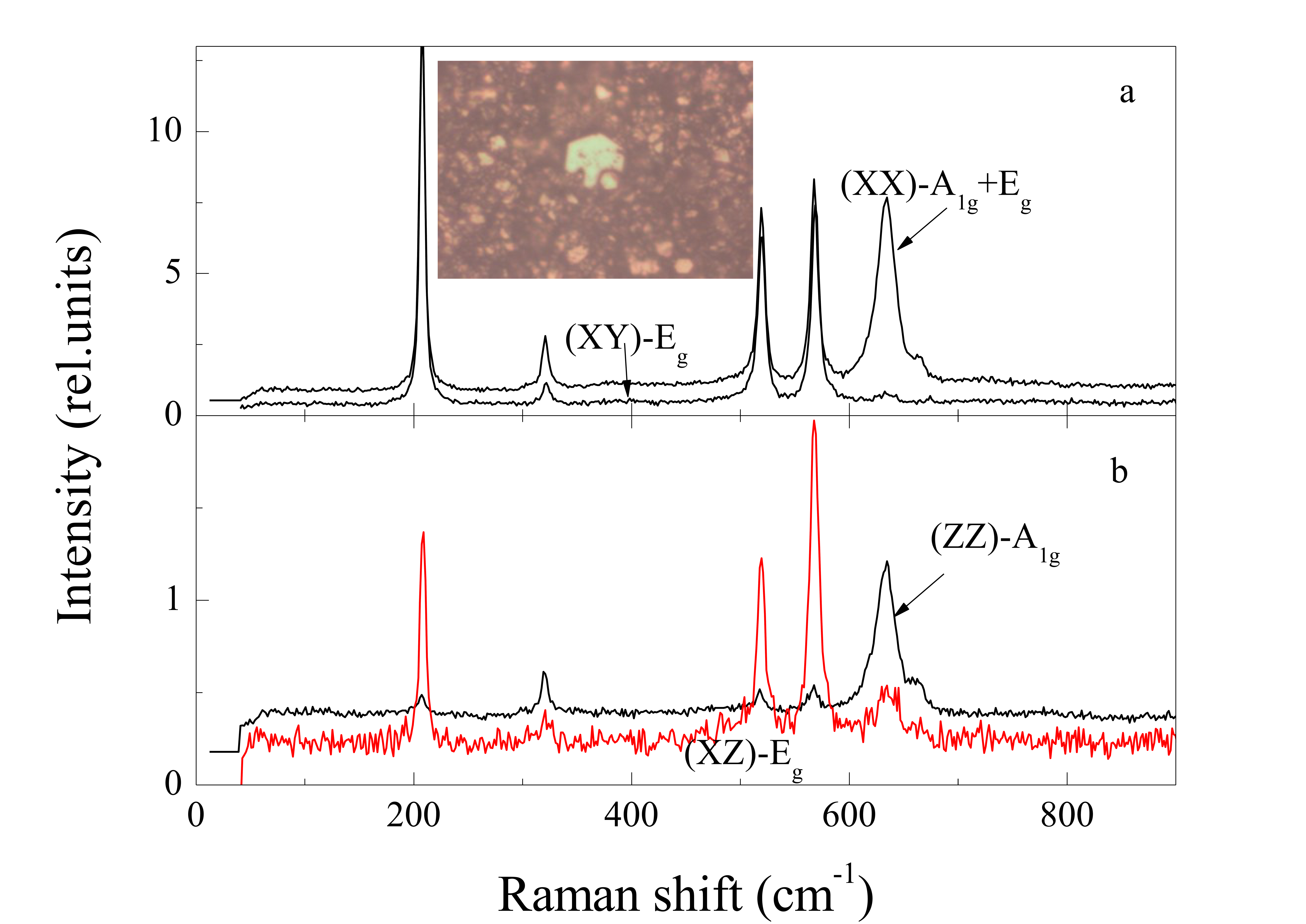}
\caption{\label{phonons} Experimental phonon Raman spectra of SrRu$_2$O$_6$ measured at 300 K in different polarization geometries.a)-XX and XY, b) ZZ and XZ. Excitation-532 nm. Inset- micrograph of crystallites in the used sample, as seen in an optical microscope.}
\end{figure} 

Calculations of the phonon spectra were performed using frozen-phonon method implemented in Phonopy \cite{Phonopy} and electronic structure was obtained with Vienna ab initio simulation package (VASP) \cite{Vasp1}.  For these purposes we used $2\times2\times2$ supercell with N$\acute{\textrm{e}}$el antiferromagnetic ordering for in-plane nearest-neighbor Ru atoms and ferromagnetic coupling between the Ru planes. The results were obtained with 700 eV plane-wave cutoff energy and $8\times8\times8$ k-mesh. First the structure was geometrically optimized until the energy difference between two ionic iterations reached 10$^{-6}$ eV/atom. The generalized gradient approximation (GGA) was used. A loop of self-consistent calculations stopped with the difference between two electronic steps became less then 10$^{-7}$ eV. 

\section{Results and discussions}
\subsection{Phonon Raman spectra of SrRu$_2$O$_6$}
For the hexagonal structure ($P\bar 31m$ space group), group theory predicts 27 $\Gamma$-point phonon modes ($1A_{1u}+4A_{2u}+2A_{1g}+2A_{2g}+5E_u+4E_g$); out of which 4 modes ($2A_{1g}+4E_g$) are Raman active. Fig.~\ref{phonons} shows the polarized Raman spectra of SrRu$_2$O$_6$ measured at  300 K in the in-plane (XX) and (XY) and out-of-plane (XZ) and (ZZ) polarizations. They probe  the $A_{1g}+E_g, E_g$, $E_g $ and $A_{1g}$ symmetry channels, respectively. At room temperature we observe all Raman active vibrations: four $E_g$ phonon modes at 206.2, 319.2, 517.8, 566.7 cm$^{-1}$ in (XX), (XY) and (XZ) polarizations and two $A_{1g}$ modes at 318.3, 632.5 cm$^{-1}$ in (XX) and (ZZ) polarizations. Another weak and rather narrow line was observed in the XX and ZZ polarization geometries at 665  cm$^{-1}$. As we shall see later, SrRu$_2$O$_6$ has a strong two-phonon spectrum. This $A_{1g}$ line most probably belongs to that part of spectrum. 

\begin{figure}[b]
\includegraphics[width=0.5\textwidth]{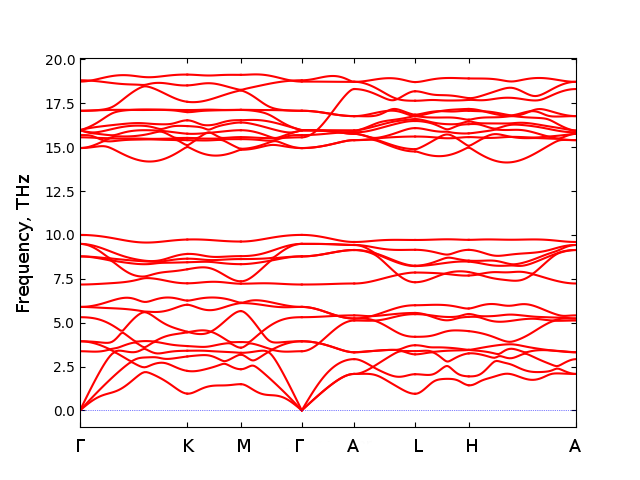}
\caption{Calculated in the GGA phonon spectrum for SrRu$_2$O$_6$.}
\label{Bands}
\end{figure} 

Tab.~\ref{freq} shows all DFT calculated phonon frequencies at the $\Gamma$ point, while in Fig.~\ref{Bands} full spectrum is presented. According to the lattice dynamic calculations frequencies of the $E_g$ modes were found at 196.6, 316.4, 533.0, and 569.8 cm$^{-1}$. The energies of the $A_{1g}$ modes were calculated at 333.6, 627.5 cm$^{-1}$. Thus, a fairly good agreement has been obtained between experiment and calculation.
 
\begin{figure}[t]
\includegraphics[width=0.4\textwidth]{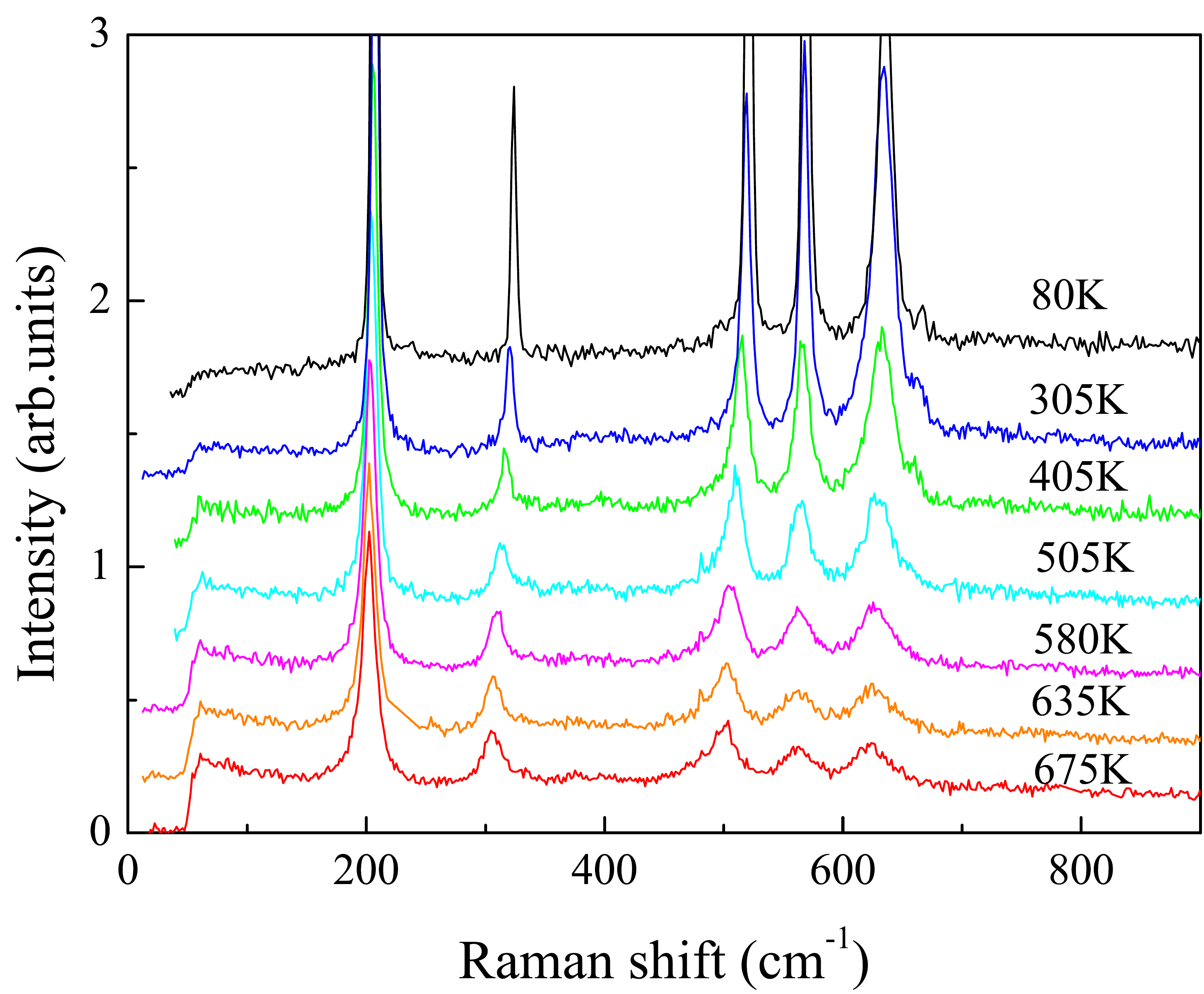}
\caption{\label{phonons-temperature} Experimental phonon Raman spectra of SrRu$_2$O$_6$ measured at different temperatures. Polarization geometry XX. Excitation - 532 nm.}
\end{figure}
\begin{figure}[b]
\includegraphics[width=0.5\textwidth]{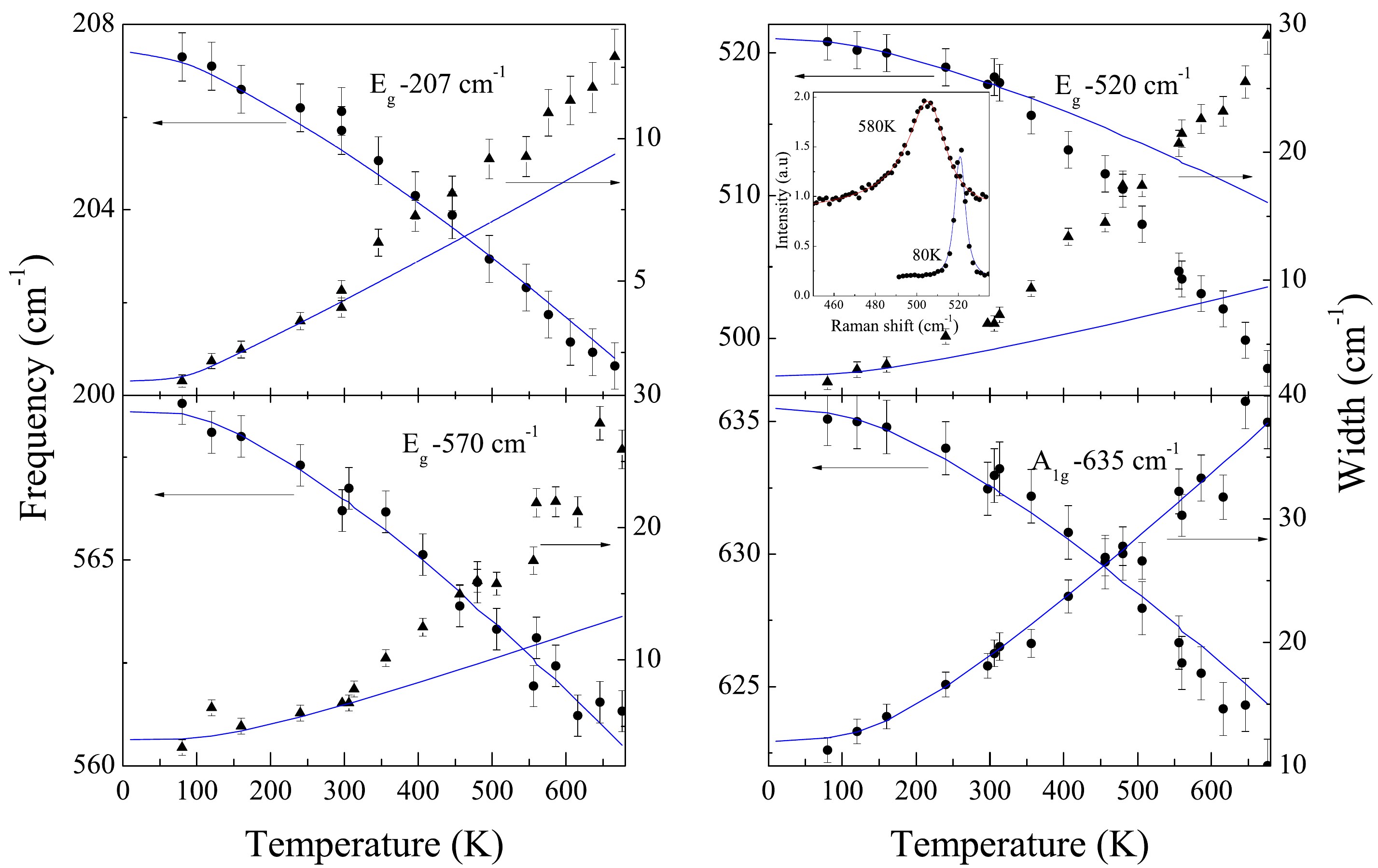}
\caption{\label{anharm} Temperature dependencies of the frequency shifts and linewidths for four Raman active phonons. Solid lines show fits to a standard anharmonic-decay model ~\cite{kl}. The inset shows the 520 cm$^{-1}$ phonon profiles measured at two temperatures with fitting by the Lorenz and Fano profiles.}
\end{figure}

Measured in (XX) polarization geometry Raman spectra are plotted in Fig.~\ref{phonons-temperature} as a function of temperature. The frequencies of all phonons soften gradually without any features with increasing temperature, and the widths increase, which can be explained by anharmonic effects. In the simplest form, the temperature behavior of phonon self-energies is described by formulas that assume the decay of an optical phonon into two phonons of half the frequency with opposite wave vectors~\cite{kl}. We fitted low-temperature dependencies of the frequency and linewidth for four of six Raman active phonons; the exact parameters of phonons close in frequency ($E_g$ and $A_{1g}$ near 320 cm$^{-1}$) are difficult to determine because of their overlap due to leakage of polarized spectra. To estimate the contribution of the thermal expansion to the phonon softening at high temperatures, we used the data from ref.~\cite{Hiley2014}; the unknown Gr\"{u}neisen coefficients were determined by fit. For high-frequency $A_{1g}$ and $E_g$ phonons at 570 and 635 cm$^{-1}$ we obtained a rather good description of the temperature dependencies of the frequencies in the whole investigated range with the Gr\"{u}neisen parameter $\gamma$ = 0.7 (Fig.~\ref{anharm}). The frequency of the $E_g$  phonon at 207 $cm^{-1}$ can be fitted with $\gamma$ = 1.5. However, for the $E_g$  phonon at 520 cm$^{-1}$, the use of an intermediate $\gamma$ = 1 does not lead to agreement with experiment. With increasing temperature, this phonon also demonstrates a huge increase in the linewidth, which is three times as large as the estimated anharmonic contribution.  Both observed effects an abnormal softening of this mode and anomalous increase in linewidth obviously indicate the existence of an additional mechanism of interaction. Another evidence of this is a significant increase in the asymmetry of the phonon profile of this phonon at high temperatures (Fig.~\ref{anharm}). If at low temperatures the phonon line is perfectly described by Lorentzian, then at high temperatures its shape was fitted by the Fano profile~\cite{klein} , which implies interference of the phonon with the continuum. The existence of additional mechanism of the phonon self-energy renormalization is confirmed by the fact that the linewidth for another three $E_g$ phonons can also not be described at high temperatures by the contributions expected from anharmonic phonon-phonon interaction (Fig.~\ref{anharm}). 

Thus, we did not observe any new phonon lines near the magnetic transition temperature. Therefore our results exclude any structural phase transition at $T_N$=565 K. However, a number of $E_g$ phonons start a gradual significant increase of linewidths and an anomalous softening at temperatures above room temperature, reaching the maximum in the paramagnetic state.  Earlier an anomalously strong growth in the linewidths (respectively, damping constants) with an increase in temperature was detected in narrow-band semiconductors FeSi and FeSb$_2$~\cite{rasu} which was explained by electron-phonon interaction.

\begin{table}[t]
 \begin{tabular}{cccc}  
  \hline \hline
No. & Mode & Calculated  & Experimental \\
       &          & frequency, cm$^{-1}$ & frequency, cm$^{-1}$ \\
 \hline
1  & A$_{1g}$ & 627.5 & 634.0 \\
2  & A$_{1u}$ & 625.3 &  IR\\
3  & E$_g$    & 569.8 & 567.7 \\
4  & E$_g$    & 533.0 & 519.3 \\
5  & E$_u$    & 532.5 &  IR \\
6  & A$_{2g}$ & 523.5 &  IR \\
7  & A$_{2u}$ & 519.0 &  IR \\
8  & E$_u$    & 498.7 &  IR \\
9  & A$_{1g}$ & 333.6 & 320.7 \\
10 & E$_g$    & 316.5 & 321.7 \\  
11 & E$_u$    & 292.8 &  IR\\
12 & A$_{2u}$ & 239.4 &  IR\\
13 & E$_g$    & 196.6 & 207.6 \\
14 & A$_{2g}$ & 177.1 &  IR\\
15 & E$_u$    & 131.4 &  IR \\
16 & A$_{2u}$ & 112.4 &  IR\\
 \hline \hline
 \end{tabular}
	\caption{\label{freq}Comparison between calculated (at the $\Gamma$ point) and experimentally found values of phonon frequencies for SrRu$_2$O$_6$. IR means that this line can be observed only by the infrared spectroscopy.}
		\label{table}
\end{table}

\subsection{Electronic Raman scattering in SrRu$_2$O$_6$.}
\begin{figure}[t]
\includegraphics[width=0.5\textwidth]{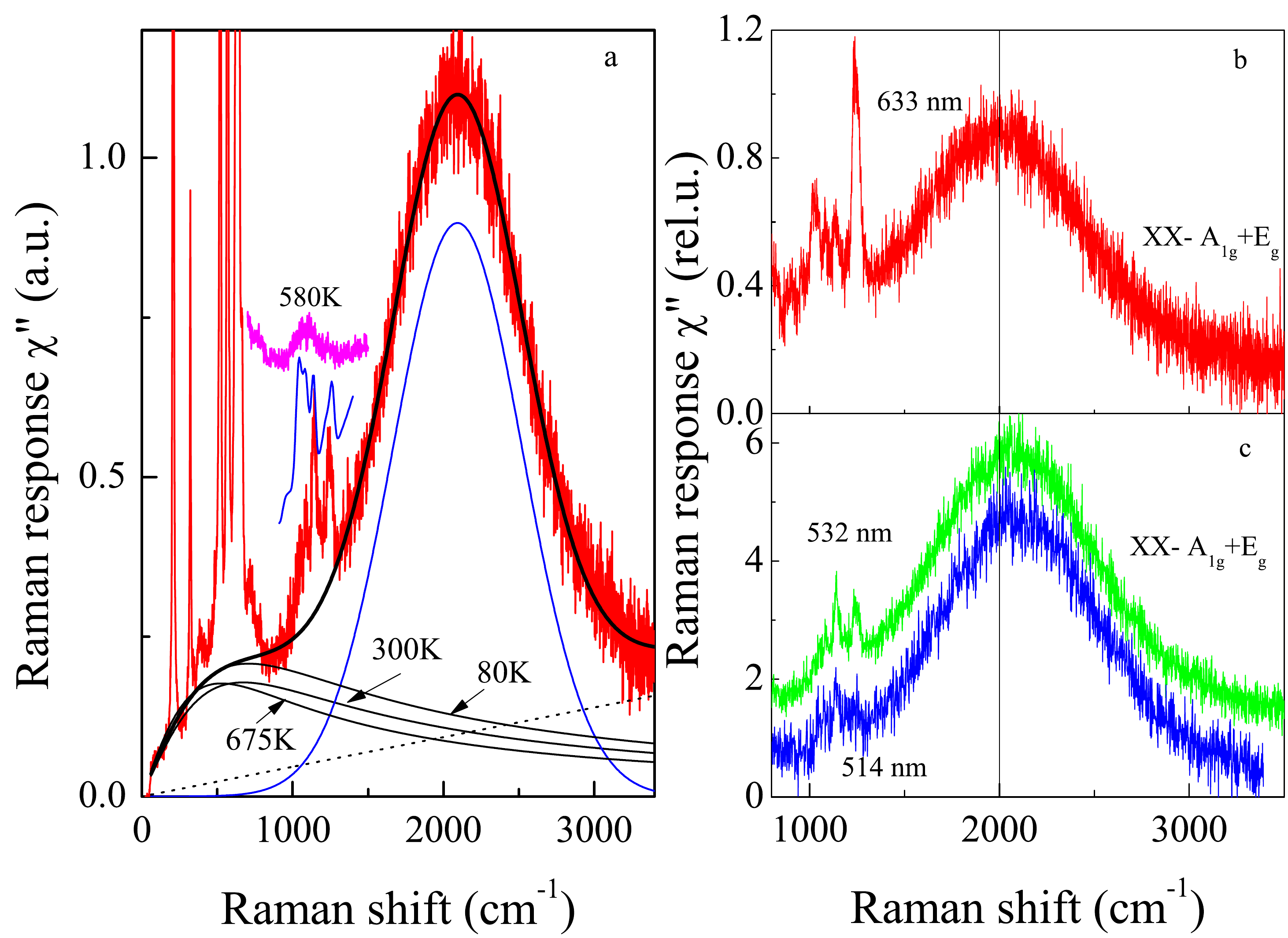}
\caption{\label{Electron-Raman-XX} a) The Raman spectra of SrRu$_2$O$_6$ measured in XX geometry at 300K. Excitation line - 532 nm. Fragments of the calculated two-phonon spectra and measured high-temperature spectra near 1200 cm$^{-1}$ also are shown. Solid lines are fits of two broad electronic bands and total profile, dotted line-linear background. Fits for the first band are shown for 3 temperatures.  b,c) Raman spectra measured with excitation by different lasers at 300K.}
\end{figure}

As can be seen in Fig.~\ref{phonons-temperature}, the intensity of the background on which phonons are superimposed increases in the low-frequency region of the spectrum as the temperature increases. Fig.~\ref{Electron-Raman-XX}(a) shows the T = 300K Raman response $\chi^{''}(\omega) = I(\omega)/(n(\omega)+1)$, (where $n(\omega) + 1$ is the Bose-Einstein factor) of SrRu$_2$O$_6$  in the spectral range from 50 to 3500 cm$^{-1}$. The low-frequency response linearly goes to zero at $\omega\rightarrow0$ and has a maximum near 600 cm$^{-1}$.  This peak was observed in all polarization geometries XX, XY, XZ, and ZZ, although the intensity of the continuum for polarizations in the XY plane is $\approx$3 times higher. This broad peak, obviously, is an electronic Raman scattering. The existence of electronic excitations in the phonon frequency range is confirmed by a previously noted asymmetry of some phonon lines and anomalous behavior of their self-energies indicating their interference with an interacting continuum. Observed continuum overlaps with another higher frequency band, so its shape is difficult to accurately describe. We used the following expression~\cite{car} with electron relaxation rate $\gamma\approx600$ cm$^{-1}$ to describe this peak (Fig.~\ref{Electron-Raman-XX}a): 
\begin{equation}
\chi^{''}(\omega)\propto N_f\frac{\omega\gamma}{\omega^2+\gamma^2}	
\end{equation}
where $N_f$ is the density of states at the Fermi level. Such a fitting shows rather weak decrease of the continuum frequency and intensity when the temperature varies from 80 to 675 K. Such weak dependence of the shape and intensity of the observed continuum on temperature is surprising.  This probably suggests that the phonons do not contribute to its thermal relaxation. A possible reason for this behavior presupposes the presence of defects that leads to the scattering of electrons near the Fermi level~\cite{car}. Consequently, the question arises of the origin of the interacting continuum, the existence of which is indicated by the observed interference with phonons and which should have a temperature-dependent density of states.

Somewhat above this peak, one can see a group of rather narrow lines (1100-1200 cm$^{-1}$), which have frequencies twice as large as group of  Raman peaks (500-600 cm$^{-1}$). Like these lines of the one-phonon spectrum, they soften and substantially broaden with increasing temperature (Fig.~\ref{Electron-Raman-XX}(a)). We believe that this structure is due to two-phonon Raman scattering, which is confirmed by our calculations of the two-phonon spectrum shown in Fig.~\ref{Electron-Raman-XX}(a).

At higher frequencies, a broad peak at $\approx$2050 cm$^{-1}$ is observed. This peak lies outside the energies of the one-phonon region and  is supposedly of an electronic nature. In order to rule out luminescence as the origin of the high-frequency broad band, Raman spectra were recorded with a different laser lines (633 and 514 nm). Observation of this peak with excitation by different lasers (Fig.~\ref{Electron-Raman-XX}b) confirms that this peak is electronic Raman scattering.  The peak frequency (fitted by Gaussian) and its width ($\approx$1000 cm$^{-1}$) increase somewhat with increasing excitation energy. The intensity of the peak increases substantially with increasing excitation energy (514 and 532 nm), which may indicate a resonance with transitions near 2.4 eV. The explanation of such a resonance dependence of the intensity is supported by calculations of the electronic structure SrRu$_2$O$_6$~\cite{Okamoto2017a}, showing that the underlying zones are located at a distance of ~ 2 eV from the Fermi level.

\begin{figure}[t]
\includegraphics[width=0.5\textwidth]{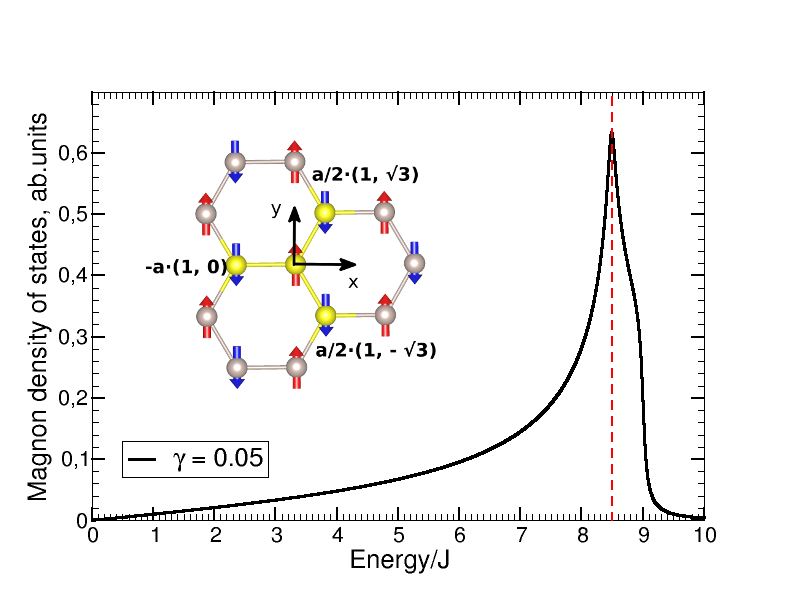}
\caption{ \label{DOS} Magnon density of states for SrRu$_2$O$_6$ obtained by integration of the (\ref{SWS_HCL}) spectrum. Inset shows the coordinate system used in the calculations.}
\end{figure}

There can be two mechanisms responsible for this peak: two-magnon scattering and interband electronic transitions. Let us start with estimation of possible energy for the two-magnon excitation.  We approximate spin-Hamiltonian of SrRu$_2$O$_6$ by the Heisenberg model (i.e. neglect, e.g., the single-ion anisotropy, which was found to be $D \sim 0.8$ meV\cite{st}). Within the linear spin-wave theory the spectrum of the N$\acute{\textrm{e}}$el's two-sublattice AFM with the Hamiltonian
\begin{equation}
\label{H}
H = - 2 \sum_{\langle ij \rangle} J_{ij}\mathbf{S}_i\mathbf{S}_j,
\end{equation}
where $J_{ij}$ are exchange parameters and summation runs over spin pairs $\mathbf{S}_i\mathbf{S}_j$, can be obtained analytically using general formalism given, e.g., in Ref.~\cite{Frederic} or ~\cite{Kittel}. If for simplicity we take into account only nearest neighbor exchange between spins of different sub-lattices, $J$, then the spin-wave dispersion can be calculated as  
\begin{equation}
\label{SWS_of_N2SLM}
\omega_{\mathbf{k}}^2 = {A_{\mathbf{k}}}^2 -|B_{\mathbf{k}}|^2.
\end{equation}
Explicit expressions for $A_{\mathbf{k}}$ and $B_{\mathbf{k}}$ can be found in Ref.~\cite{Frederic}. In the case of the honeycomb lattice with a N$\acute{\textrm{e}}$el AFM ground state, $S = 3/2$, and coordinate system chosen as shown in the inset of Fig.~\ref{DOS} the spin-wave dispersion is given by:
\begin{eqnarray}
\label{SWS_HCL}
\omega_{\mathbf{k}}^2 &=& 81J^2 - 9J^2 \Bigl|\exp \left(\frac{a}{2}k_x+\frac{a \sqrt{3}}{2}k_y\right) \nonumber \\
&+& \exp \left(\frac{a}{2}k_x-\frac{a \sqrt{3}}{2}k_y \right) + \exp \left(-a k_x \right) \Bigl|^2 ,
\end{eqnarray}
where $a$ is the Ru-Ru distance.

By integration \eqref{SWS_HCL} over the first Brillouin zone with the reciprocal-lattice vectors $\mathbf{b}_1 = \frac{2 \pi}{3a} (1, \sqrt{3})$ and $\mathbf{b}_2 = \frac{2 \pi}{3a} (1, \sqrt{3})$ one can get the magnon density of states (DOS). This integration was performed by direct summation over 250000 $\mathbf{k}-$points in the first Brillouin zone with the Lorentzian having width $\gamma = 0.05$. 

The results of these calculations are shown in Fig.~\ref{DOS}. There is a strong peak at $8.5J$ associated with the van Hove singularity in spin-wave spectrum and a shoulder at $9J$. While accurate computation of two-magnon excitations requires calculations of vertex corrections \cite{per}, one may roughly estimate the energy of two-magnon processes from magnon DOS. Thus, one may expect that the frequency of two-magnon excitation will be of order of $17J$. 

\begin{figure}[t]
\includegraphics[width=0.5\textwidth]{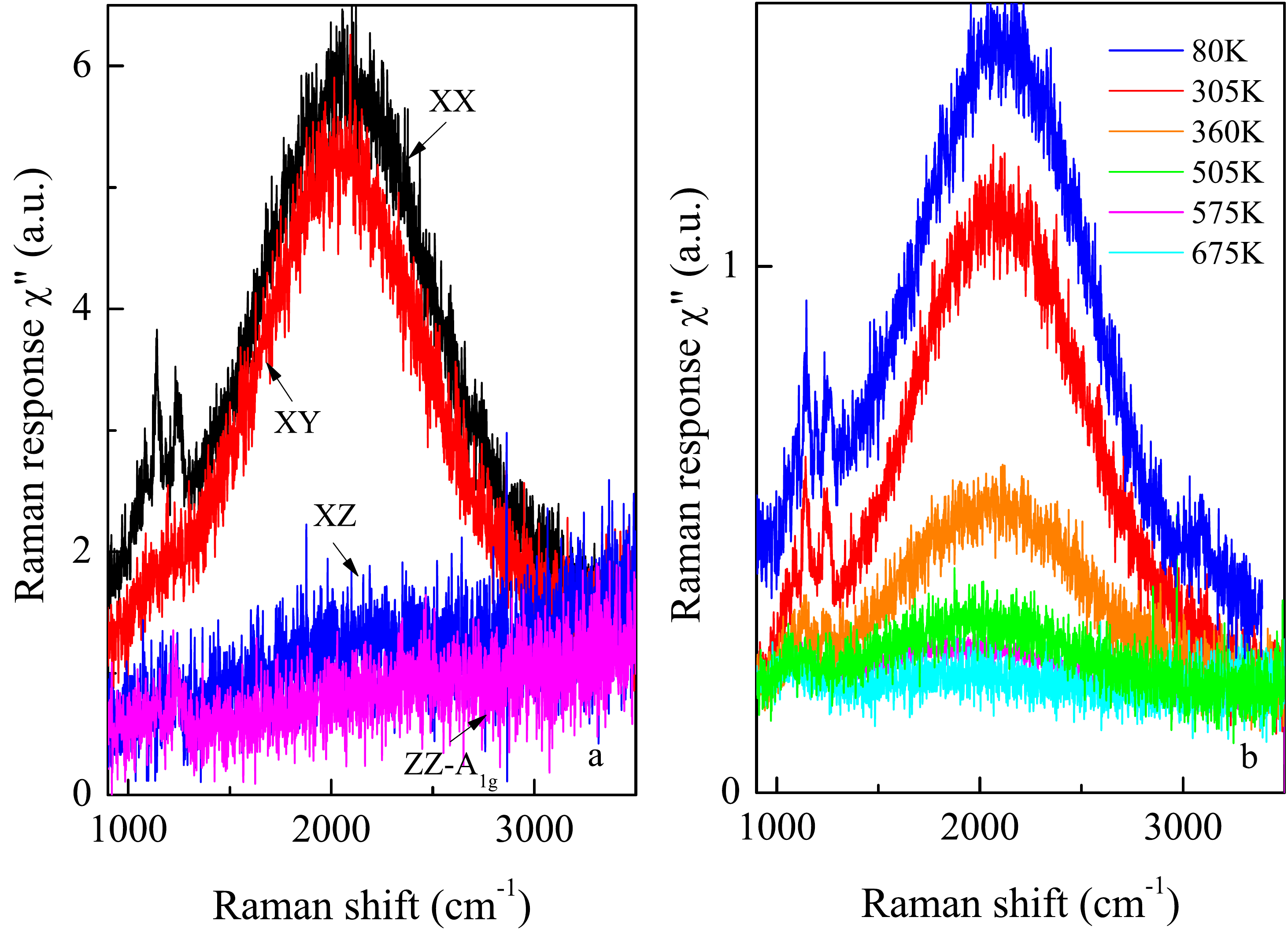}
\caption{\label{Electronic-raman-diffpol} a) High-frequency Raman spectra of SrRu$_2$O$_6$ measured at 300K in different polarization geometries. b) High-frequency Raman spectra of SrRu$_2$O$_6$ measured at different temperatures.}
\end{figure}

Two estimations of the exchange integral in the $ab-$plane using previous DFT results can be made. First of all, $J$ was calculated in the localized electron model for $S=3/2$ Heisenberg hamiltonian using hopping parameters extracted from the GGA and interaction matrix computed within constrained random-phase approximation (cRPA)~\cite{Tian2015}. Taking into account additional factor of two in our model \eqref{H} we get $J \sim 600$~K. Alternatively, $J$ can be recalculated from Eq. (1) of Ref.~\cite{st}, which was directly obtained from the GGA total energies and thus takes into account strong suppression of the magnetic moment down to 1.3 $\mu_B$ in SrRu$_2$O$_6$. Fitting Eq. (1) in Ref.~\cite{st} by \eqref{H} in the present work (which is written for $S=3/2$, not for magnetization as Eq. (2) in Ref.~\cite{st}) one gets $J \sim 500$~K.  These two estimations translate to frequencies of two-magnon excitations $\sim$6000-7000 cm$^{-1}$, which are much higher than observed peak at 2050 cm$^{-1}$.  Although an account of the final state magnon-magnon interaction can slightly shift the peak down, we feel that either DFT calculations overestimate exchange parameters (and hence magnon energies) or the nature of this peak is not two-magnon scattering\footnote{We note also that in real SrRu$_2$O$_6$ magnetic moment is strongly suppressed with respect to ionic $S=3/2$. Our estimation of the exchange constant based on the GGA total energies \cite{st} takes this effect into account. Crude semiclassical estimation of this effect for $J$ calculated in Ref.~\cite{Tian2015} reduces energy of the two-magnon excitation in two times, but even in this case it is much higher than peak position in the experiment.}. Also temperature dependence of the intensity of the 2050 cm$^{-1}$ peak is not characteristic for two-magnon scattering of light, the intensity of which typically remains high at $T>T_N$, since short-range magnetic correlations persist even above Neel temperature. The intensity of this peak, however, decreases substantially (by 5 times) in the temperature range 300-500 K and becomes negligible in the region $T_N$ (Fig.~\ref{Electronic-raman-diffpol}b). Also, the frequency and width of this peak only slightly decreases and is broadened (both by 5-10\%) with an increase in the temperature to $T_N$. 

A broad, quasicontinuous polarization-independent Raman response has been predicted for the Kitaev spin liquid ~\cite{kn}. The presence of such a features near 3000 cm$^{-1}$ in  magnetically ordered iridates ~\cite{gu,gu1} was interpreted as evidence for proximity to the Kitaev spin liquid. It is difficult to assume the existence of a quantum spin state in SrRu$_2$O$_6$. In addition, the observed electronic excitation near 2050 cm$^{-1}$ have a clear symmetry, suggesting its connection with the honeycomb layer.

Another possibility is an electronic transition through the band gap. One may see from Fig.~\ref{Electronic-raman-diffpol}a that the 2050 cm$^{-1}$ peak is observed with the same intensity in the XX, XY polarizations and is absent in the XZ and ZZ polarizations. For the crystal structure of SrRu$_2$O$_6$  with $P\bar31m$  space group and $D_{3d}$ point group this implies the $E_g$ symmetry of the observed peak. However,  there must be the XZ component in the $E_g$ Raman tensor. This means that the transitions must be described by a higher symmetry than crystal structure of SrRu$_2$O$_6$ provides. The minimal supergroups of $P\bar31m$ are $P6/mmm$ and $P63/mcm$ and the point group in both cases is $D_{6h}$. Only in $D_{6h}$ point group the $E_{2g}$ Raman tensor contains just XX and XY components and thus we conclude that the excitations resulting in the 2050 cm$^{-1}$ peak in Raman spectrum should be described not by $D_{3d}$, but rather by a more symmetric $D_{6h}$ point group. 

This is exactly the point group of benzene molecule, which in particular describes symmetry of the molecular-orbitals proposed in Ref.~\cite{st}. This, however, does not automatically justifies molecular-orbital picture. Within the $D_{6h}$ point group, the symmetry of 2050 cm$^{-1}$ peak is $E_{2g}$. Among all symmetry allowed transitions we are interested in case of SrRu$_2$O$_6$ only in the following: $A_{1g} \leftrightarrow E_{2g}$, $B_{1u} \leftrightarrow E_{1u}$, $A_{1g} \leftrightarrow A_{1g}$, $B_{1u} \leftrightarrow B_{1u}$,  $E_{2g} \leftrightarrow E_{2g}$, $E_{1u} \leftrightarrow E_{1u}$. 

One may see, that there are symmetry allowed transitions in both theoretical models proposed to explain physical properties of SrRu$_2$O$_6$. In the picture of localized correlated electrons these  will be transitions between trigonal $e_g$ and $a_{1g}$ orbitals or within these manifolds. It is easy to estimate corresponding excitation energies, using spectral function plots presented in Ref.~\cite{Okamoto2017a}  for magnetically ordered phase: $E \sim 0.5$ eV, i.e. it is about factor of two larger than experimental value.

It is much more complicated to estimate excitation energy across the band gap in the molecular-orbital based picture. One might expect that the energy of the first electron transition should be 0.4 eV (first peak above the Fermi level in Fig. 2 of Ref.~\cite{Pchelkina2016}). At first sight this transition is forbidden by symmetry, since this would be the transition between the $E_{2g}$ and $E_{1u}$ molecular-orbitals, but stabilization of the long range (N$\acute{\textrm{e}}$el AFM) magnetic order results in a mixing of the $B_{1u}$ and $E_{2g}$ states. Moreover, there is spontaneous degeneracy between the $E_{1u}$ and $A_{1g}$ molecular-orbitals in SrRu$_2$O$_6$~\cite{st}. Second, the excitation energy should be  definitely smaller than 0.4 eV obtained from the GGA calculations. It is nearly impossible to take into account correlation effects in molecular-orbital model in the DFT+DMFT calculations (since this would imply solution of the impurity problem with $6 \times 3 \times 2 = 36$ spin-orbitals; an impurity is the Ru hexagon), but qualitative effect of the electronic correlations in covalent insulator\cite{Hariki2017} is very well known - they result in narrowing of the band gap~\cite{Kunes2008a}. In FeSb$_2$ having rather similar bandwidths this leads to a decrease of the band gap in two times for Hubbard $U=1.5$ eV~\cite{Kunes2008a}. Thus, one may expect that an account of correlation effects would give the excitation energy of $\sim$0.2-0.3 eV close to what we obtained in the experiment. The observed anomalous behavior of phonon self- energies is an evidence in favor of such a scenario.

\begin{figure}[t]
\includegraphics[width=0.4\textwidth]{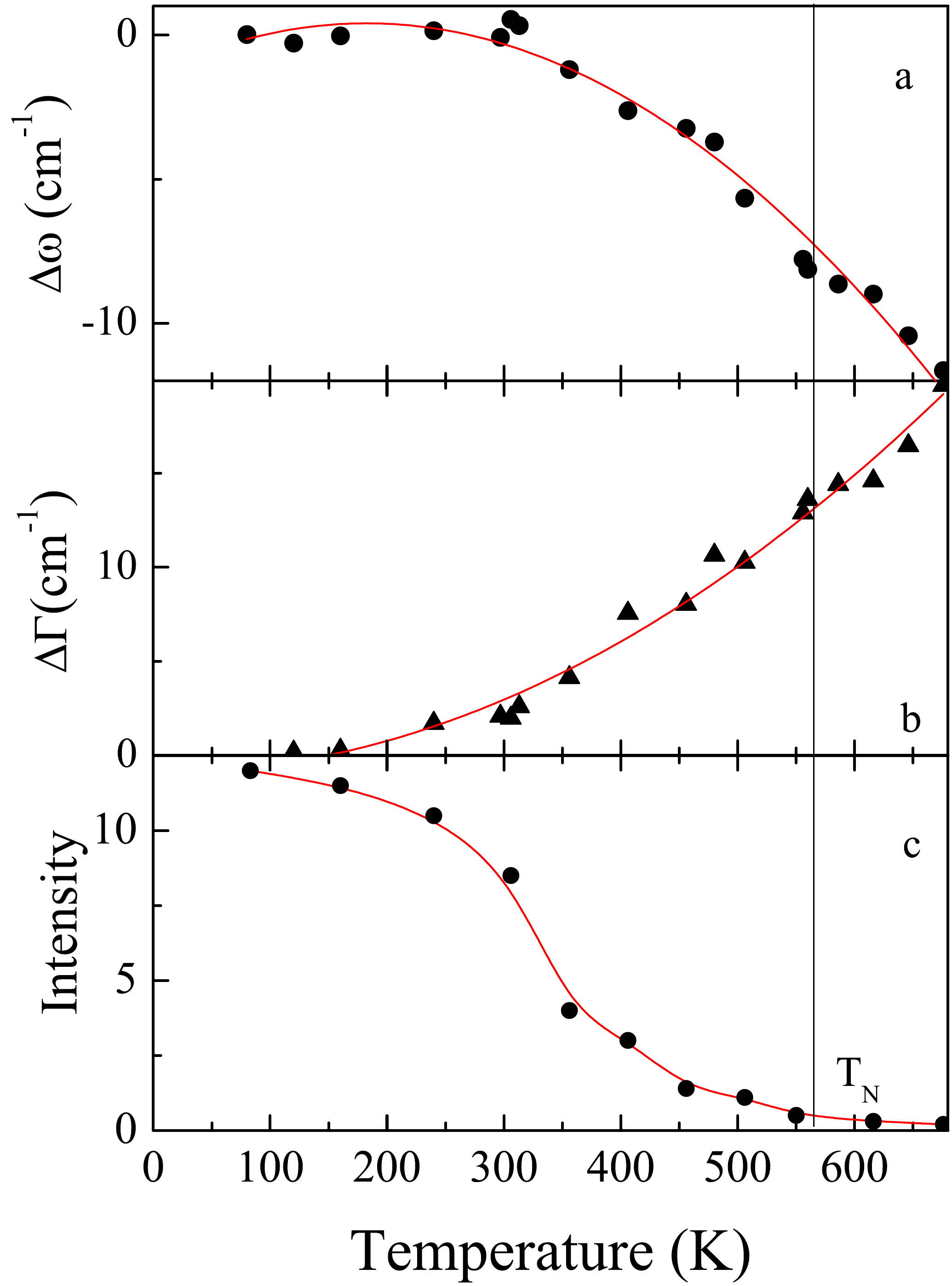}
\caption{\label{fr-damp-2050} Temperature dependencies of a) extra  520 cm$^{-1}$ phonon softening ($\Delta\omega$=difference between measured frequency and its anharmonic fit in Fig.~\ref{anharm}), b) extra 520 cm$^{-1}$ phonon damping ($\Delta\Gamma$=difference between measured width and its anharmonic fit in Fig.~\ref{anharm}) and c) intensity of the 2050 cm$^{-1}$ broad band in  SrRu$_2$O$_6$. Solid lines are guides to the eye.}
\end{figure}

Fig.~\ref{fr-damp-2050} shows the temperature dependencies of the excess softening and broadening of the phonon line at 520 cm$^{-1}$, which are obviously due to the interaction with electrons. These excess phonon self energies indicate an increase in the density of states of electronic excitations interacting with phonons. As can be seen from Fig. ~\ref{fr-damp-2050}, an increase in the density of these excitations at phonon frequencies correlates with a decrease in the intensity of the band at 2050 cm$^{-1}$. It should be noted that in the same temperature range, an increase in the magnetic susceptibility is observed at temperatures below and above $T_N$~\cite{Hiley2015}, which is probably due to the development of spin fluctuations. It can be assumed that the spin-phonon interaction is responsible both for the temperature-dependent excess phonon self-energies and for the effect on the structure of transitions between molecular orbitals, although this issue requires further study.

\section{Conclusions}
Inelastic scattering of light by electronic and phonon excitations has been investigated in SrRu$_2$O$_6$, analysis and comparison with calculations have been carried out. The temperature dependences of the frequencies and widths of some $E_g$ phonons are not  characteristic of the behavior of anharmonic contributions and show gradual anomalous extra frequency softening and linewidth broadening in the transition to the paramagnetic state. We also do not observe any additional phonon modes at the magnetic transition, thus ruling out any possibility of a structural transition. 

In addition to phonon lines we detected broad bands in the low-frequency (600 cm$^{-1}$) and high-frequency regions of the spectrum (2050 cm$^{-1}$). The peak at 2050 cm$^{-1}$ can potentially be explained by two-magnon processes, but available estimations of the exchange parameters for the Heisenberg model obtained by the density functional calculations are inconsistent with the peak position. Thus, we associate this peak with electronic scattering. These excitations have a clear $E_{2g}$ symmetry and can be associated with transitions between molecular orbitals, but further theoretical calculations are needed to estimate role of electronic correlation effects.

\section*{Acknowledgements}
We are grateful to J. Buhot, I. Mazin, G. Khaliullin and N. Perkins for useful discussions. This research was carried out within the state assignment of FASO of Russia (No. AAAA-A18-118020290104-2, No. AAAA-A18-118020190098-5). The research was partially supported by the grant of the Russian Foundation for Basic Research (project no. 16-02-00451), by the Russian president council on science (through MD-916.2017.2), by the Russian Ministry of Science and High 685 Education (02.A03.21.0006), and by the Ural branch of Russian academy of science (18-10-2-37).

\end{document}